\documentclass[12pt,showpacs,showkeys,preprintnumbers,nofootinbib]{revtex4-2}

\usepackage{amssymb,amsmath,graphics,graphicx}

\begin{document}

\newcommand{\pderiv}[2]{\frac{\partial #1}{\partial #2}}
\newcommand{\deriv}[2]{\frac{d #1}{d #2}}

\title{Recent violent political extremist events in Brazil and epidemic modeling: The role of a SIS-like model on the understanding of spreading and control of radicalism}

\author{Nuno Crokidakis}
\thanks{E-mail: nunocrokidakis@id.uff.br}

\affiliation{Instituto de F\'{\i}sica, Universidade Federal Fluminense, Niter\'oi, Rio de Janeiro, Brazil 
}

\date{\today}

\begin{abstract}
\noindent
In this work we study a simple mathematical model to analyze the emergence and control of radicalization phenomena, motivated by the recent far-right extremist events in Brazil, occurred in January 8, 2023. For this purpose, we considered a compartmental SIS-like model that takes into account only the right electors, for simplicity. The model considers radical and moderated right electors, and the transitions between the two compartments are ruled by probabilities, taking into account pairwise social interactions and the important influence of social media through the dissemination of fake news. The role of the Brazilian Federal Supreme Court on the control of such violent activities is also considered in a simple way. The analytical and numerical results show that the influence of social media is essential for the spreading and prevalence of radicalism in the population. In the presence of such social media, we show that radicalism can be controlled, but not extincted, by an external influence, that models the acting of the Federal Supreme Court over the violent activities of radicals. If the social media effect is absent, the radicalism can disappear of the population, and this phenomenon is associated with an active-absorbing nonequilibrium phase transition, like the one that occurs in the standard SIS model.

\end{abstract}


\maketitle

\section{Introduction}

\qquad Radicalization phenomena occur in many countries. As a complex emergent phenomenon, it can be analyzed through the tolls of Statistical Physics \cite{social_rmp}. Indeed, such tolls were recently applied by physicists to a wide range of social phenomena \cite{galam_frontiers} like opinion dynamics \cite{sznajd,galam_review,nuno_pla_2014,nuno_andre_marcelo_pre}, alcoholism \cite{nuno_lucas1,nuno_lucas2,jin}, tax evasion \cite{nuno2022}, adoption of innovations \cite{iglesias}, rumor spreading \cite{liqing}, and many others.

Taking into account radicalism in politics, a recent work \cite{Van_Hiel} discussed that radicals are primarily driven by anger, more than by anxiety, meaning that their information processing is heavily focused on consistency and closure. Considering moderated individuals, it was also discussed that the abyss between moderates and radicals concerns whether or not to be in the political system at all. The authors in \cite{Van_Hiel} finally concluded that issue-position polarization is a phenomenon that operates to an equal extent in moderate voters than in adherents of radical and populist parties. 

Radicalism usually lead to violent activities. As pointed in a recent work \cite{kleinfeld}, recent alterations to violent groups in the United States and to the composition of the two main political parties have created a latent force for violence that can be: triggered by a variety of social events that touch on a number of interrelated identities; or purposefully ignited for partisan political purposes \cite{kleinfeld}.

Recently, in scenes resembling the events of January 6, 2021, in the US Capitol, masses of Brazilian supporters of defeated far-right former president Jair Bolsonaro are seen breaking into the country’s Congress building, presidential palace and Supreme Court. In social media videos, many demonstrators are seen breaching the buildings, climbing on top of the Congress roof and breaking the glass in its windows, also vandalizing objects inside \cite{times_of_israel}. After such events, the Renew Europe Group in the European Parliament condemns in the strongest terms the criminal actions perpetrated by such radical far-right supporters against the Brazilian Executive, Legislative and the Supreme Court and calls on the Commission to strictly enforce the Digital Services Act to fight disinformation and hatred, and to propose new legislation to curb fascism and extremism online \cite{renew_group}.

Two months after such extremist attacks to democratic institutions in Brazil, as concerns grow over the possibility of new riots in the near future, the country's new administration faces the significant challenge of countering the ongoing rise in radicalization spurred by social media \cite{forbes}. Indeed, after such attacks it was discovered the tactical use of social media to mobilize and fundraise. When Bolsonaro was originally elected in 2018, the social media platforms were already a medium for promoting disinformation, conspiracy, and fear. But in the lead-up to the violence of 8 January, they became a means by which far-right radicals raised funding for an attack and mobilized supporters to come to Brasilia (the capital of Brazil)  - advertising the availability of buses and even free food for marchers \cite{chathamhouse}. It suggests that the social media effects are fundamental to understand the emergence and spreading of radicalism. Recent works discussed the relation between extremism and social media \cite{amit,asif,kenyon}.

To counter radicalization, 24 hours after the extremist attacks, president Lula met with the 27 governors, the presidents of the House and Senate, members of the Supreme Court and the Prosecutor-General's office at the Planalto Palace in Brasilia and stated that the Brazilian institutions will investigate and locate all those who financed the extremist invasions \cite{brasil_de_fato}. In addition,
the Supreme Court minister Alexandre de Moraes has ordered three new investigations into the January 8 riots in Brazil. The inquiries will separately handle the vandalism's financial backers, instigators and perpetrators who were not arrested in the act \cite{brazilian_report}. Brazil’s Supreme Court has also agreed to investigate whether former president Bolsonaro incited the far-right mob that ransacked the country’s Congress, top court and presidential offices, a swift escalation in the probe that shows the ex-leader could face legal consequences for an extremist movement he helped build \cite{france24}.

To deal with radicalization phenomena, some mathematical models were proposed \cite{santoprete,santoprete2,nguyen,javarone_galam,galam_radicals_2022}. The works \cite{santoprete,santoprete2} considered compartmental models in order to counter violent extremism (for a recent review on the use of deterministic epidemiological models in modeling social contagion phenomena, see \cite{sooknanan}). Based on bounded-confidence models, the authors in \cite{nguyen} showed that if individuals make biased compromises, extremism may still arise without a need of an explicit classification of extremists and their associated characteristics. The phenomenon of radicalization was also investigated in the work \cite{javarone_galam}, where two subpopulations were considered, the core and sensitive ones. The results pointed out that the core (inflexible) agents have a fundamental role in hinding radicalization within the sensitive subpopulation. Finally, the authors in \cite{galam_radicals_2022} study the conditions of propagation of an initial emergent practice qualified as extremist within a population adept at a practice perceived as moderate, whether political, societal or religious. The results suggest that only a change in attitude from the moderated individuals can put a stop to the dynamics of radicalization \cite{galam_radicals_2022}.

This work is organized as follows. In section 2 we introduce the mathematical formulation of the model. In section 3 we present our analytical and numerical results, as well as the discussion of such results. Finally, in section 4 we present our final remarks.


\section{Model}

\qquad We consider a population of $N$ individuals. Since the recent extremist events in Brazil were conducted by far-right radicals, we considered for simplicity only the right electors. For this purpose, we considered only two subpopulations of such right electors: the radical ($R$) and the moderated electors ($M$). The radicals are extremist individuals that can destroy public buildings, fight with other ideologies (with far-left individuals, in this case), and try to make moderated individuals to adopt an extremist viewpoint \cite{javarone_galam,galam_radicals_2022,radical_nuno_2023}. On the other hand, moderated individuals are citizens that cannot change side, i.e., they will always vote for right candidates in the present case. However, they are not currently radicals, i.e., they do not fight with individuals with a contrary political ideology, neither are enthusiastic of the destruction of public property. Those individuals can become radicals due to some social mechanisms, that will be discussed in the following.

For simplicity, we are considering that sensitive/conformists individuals, i.e., individuals that can change from right to left, do not interact with the above-mentioned two classes, $R$ and $M$. Thus, in such a case, we only consider the transitions between the compartments $R$ and $M$. For study how an individual adopts a radical viewpoint, it appears to be realistic even though it is a very simplified view \cite{galam_radicals_2022}.

The possible transitions are given squematically as:
\begin{eqnarray} \label{eq1}
R + M  \stackrel{\gamma}{\rightarrow} & R + R ~,  \\ \label{eq2}
R  \stackrel{\epsilon}{\rightarrow} & M ~, \\ \label{eq3}
M  \stackrel{\delta}{\rightarrow} & R ~.
\end{eqnarray}
Considering a fully-connected population, we can write the rate equations based on the possible transitions given by Eqs. \eqref{eq1} - \eqref{eq3}. Thus, the following system of differential equations governs the dynamics of the population:
\begin{eqnarray} \label{eq4}
\frac{d}{dt}\,R(t) & = & \gamma\,M(t)\,R(t) - \epsilon\,R(t) + \delta\,M(t)  \\  \label{eq5}
\frac{d}{dt}\,M(t) & = & -\gamma\,M(t)\,R(t) + \epsilon\,R(t) - \delta\,M(t)  
\end{eqnarray}
\noindent
together with the normalization condition, namely
\begin{equation}\label{eq6}
R(t) + M(t) = 1 ~,
\end{equation}
\noindent
since the considered population is fixed.

Let us elaborate upon the model's parameters. The parameter $\gamma$ denotes the contagion probability that represents the conviction power of radicals ($R$) over moderated individuals ($M$). On other words, it is the rate per unit time of encounters where moderated agents become radical ones. On the other hand, the parameter $\delta$ models the impact of social media's messages (Twitter, Facebook, Whatsapp and so on) on the perception of individuals about extremism. Many of such messages can be fake news. Thus, there is a probability per unit time that a moderated individual $M$, after see messages on the mentioned social networks, decide himself that is it a good idea to become a radical individual. Finally, the parameter $\epsilon$ represents the deradicalization of radical individuals. It takes into account the action of the Federal Supreme Court over the radical action of individuals. Such action, in the extremist attacks of January 8, 2023, led to the arrest of at least 1200 individuals \cite{prison}. Thus, the parameter $\epsilon$ intends to model the fear of some radical individuals to be arrested, and in such a case they avoid to destroy public patrimony and to break into the national congress, or even to participate on violent activities.

Looking for the transitions between the compartments $M$ and $R$, specially for Eqs. \eqref{eq1} and \eqref{eq2}, one can recognize the SIS epidemic model if we identify $M\to S$ and $R\to I$, i.e., if we identify the moderated individuals $M$ as the Susceptible $S$ individuals in the SIS model, and the radical individuals $R$ as the Infected $I$ ones in the SIS model. However, Eq. \eqref{eq3} does not have a relation with diseases spread by contacts like the ones modeled by the SIS model. Indeed, Eq. \eqref{eq3} can model a spontaneous infections, without the presence of infected contacts, which does not make sense for a real epidemic. However, for the present system we are modeling, it appears to be a realistic situation, as we discussed in the previous paragraph.

In the next section we will discuss the model's results.


\section{Results}

\begin{figure}[t]
\begin{center}
\vspace{6mm}
\includegraphics[width=0.4\textwidth,angle=0]{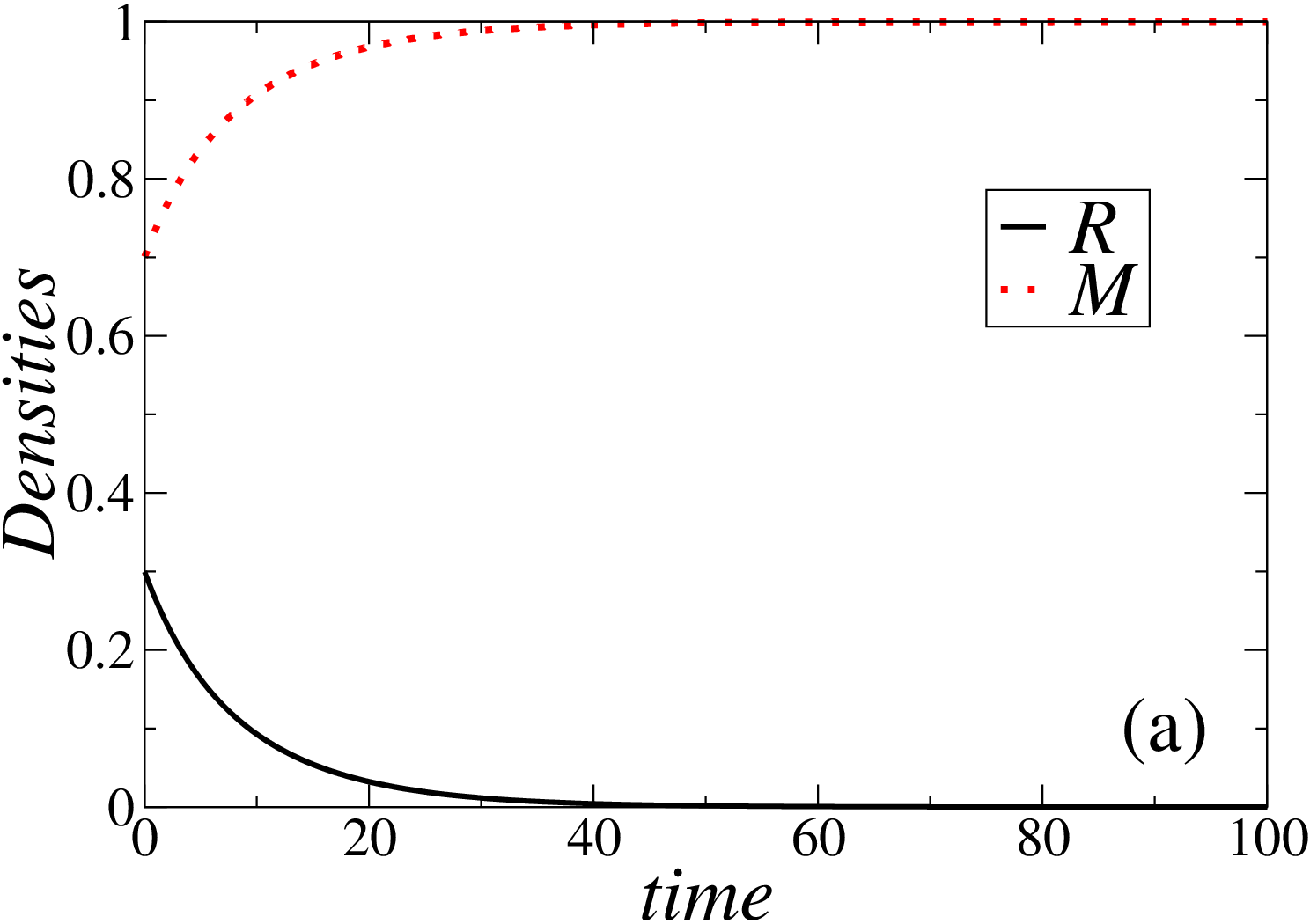}
\hspace{0.2cm}
\includegraphics[width=0.4\textwidth,angle=0]{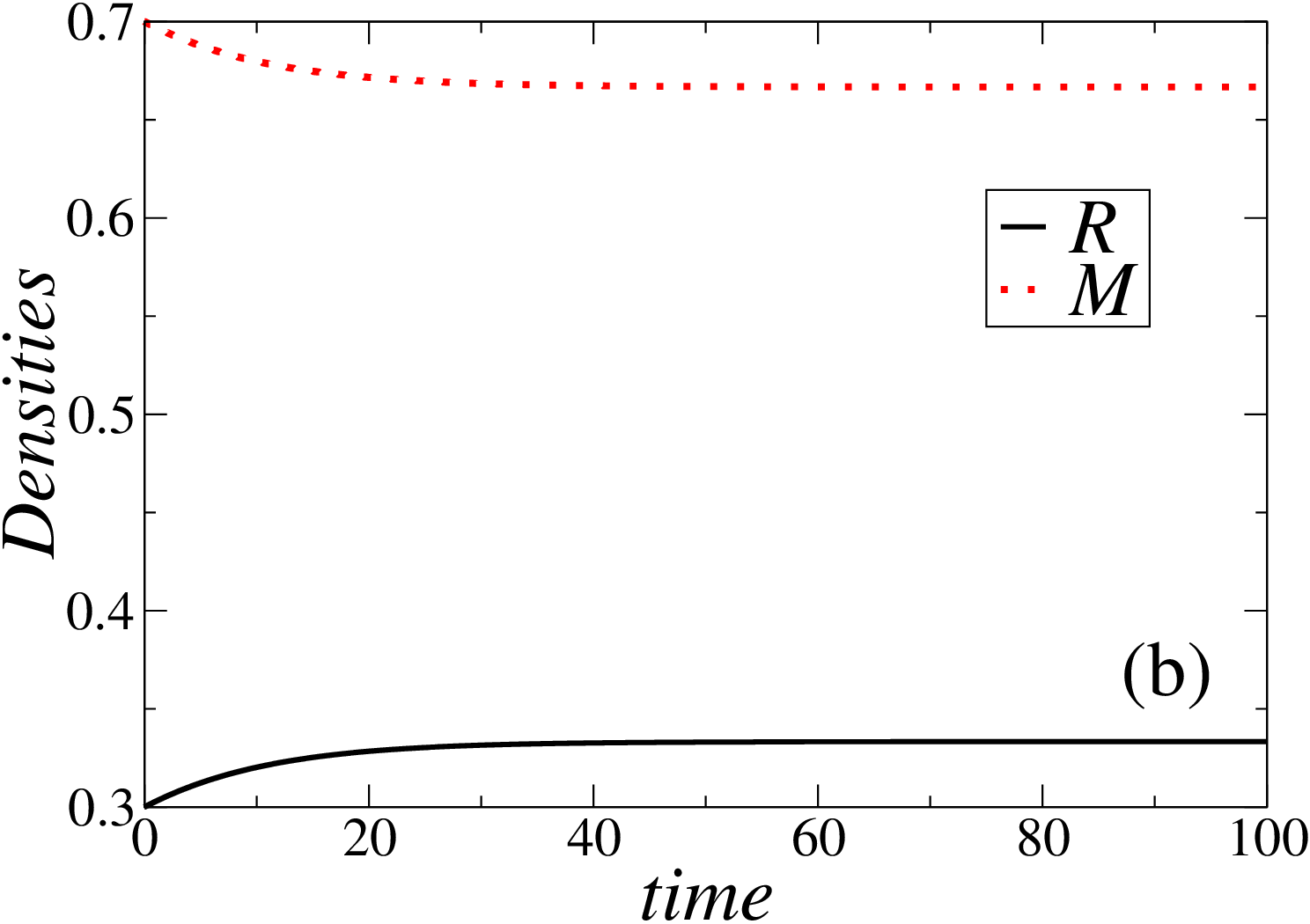}
\\
\vspace{1.0cm}
\includegraphics[width=0.4\textwidth,angle=0]{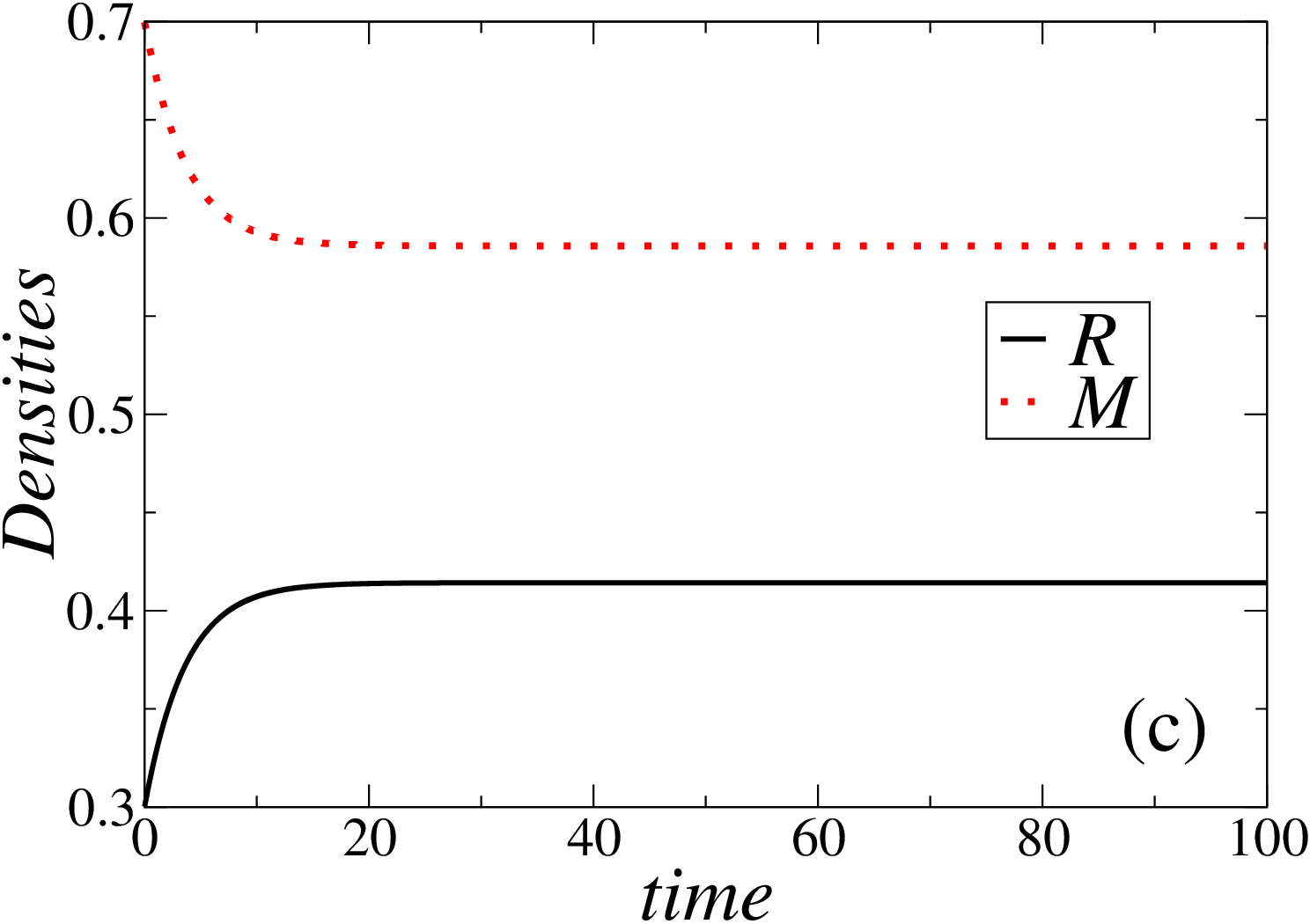}
\hspace{0.2cm}
\includegraphics[width=0.4\textwidth,angle=0]{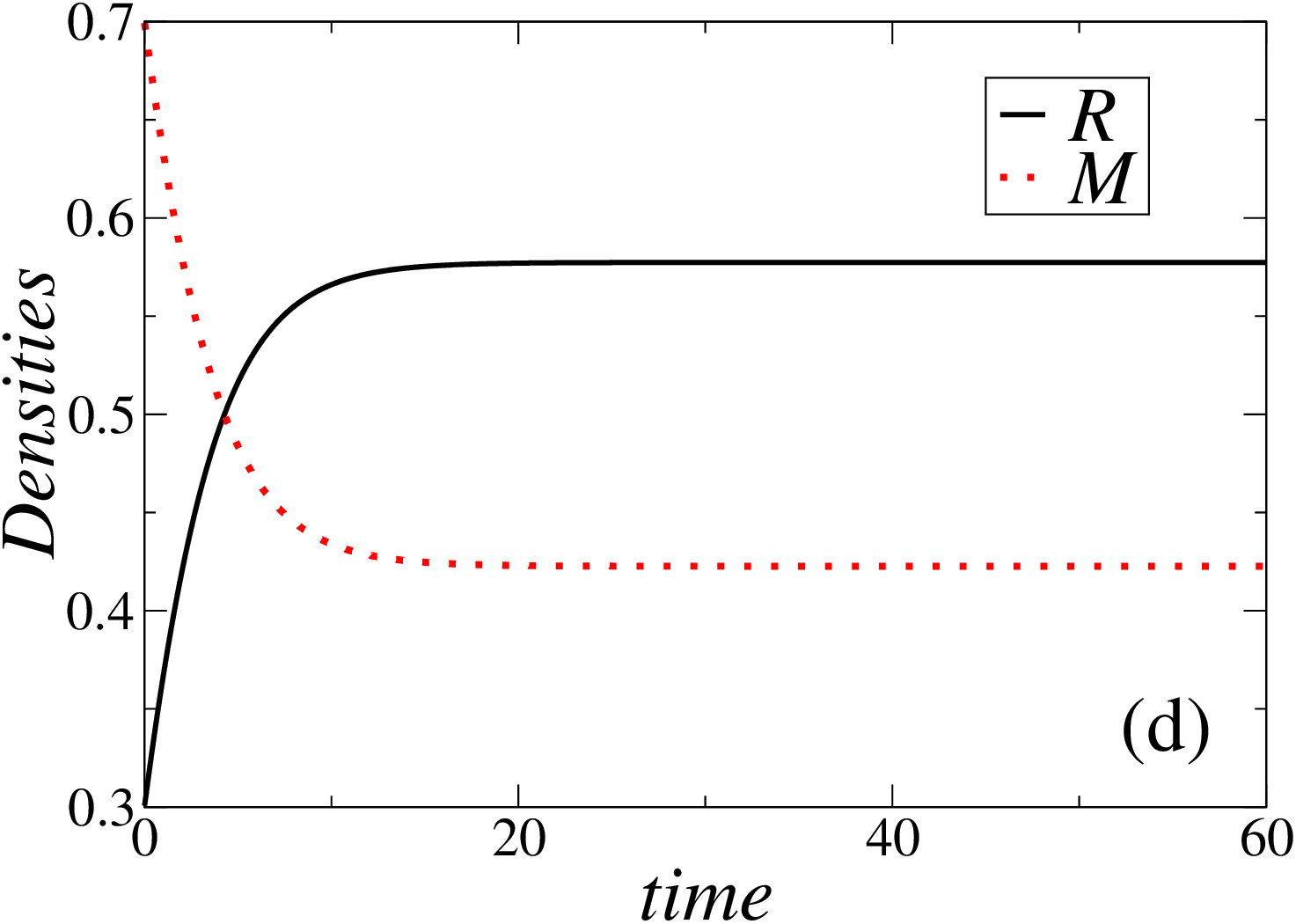}
\end{center}
\caption{(Color online) Time evolution of the densities of radical ($R$) and moderated ($M$) individuals for fixed $\epsilon=0.2$. The remaining parameters are:  (a) $\delta=0.0, \gamma=0.1$; (b) $\delta=0.0, \gamma=0.3$; (c) $\delta=0.1, \gamma=0.1$  (d) $\delta=0.1, \gamma=0.3$.}
\label{fig1}
\end{figure}

\qquad After the above discussion about the parameters, one can start analyzing the time evolution of the compartments $M$ and $R$, for typical values of the parameters. In Fig. \ref{fig1} we exhibit the time evolution of the subpopulations $R$ and $M$ for fixed $\epsilon=0.2$ and typical values of $\gamma$ and $\delta$.  The results were obtained by the numerical integration of Eqs. \eqref{eq4} and \eqref{eq5} for initial conditions $M(0)=0.7$ and $R(0)=0.3$. First, in panels (a) and (b) we show results for $\delta=0.0$, i.e., with no social media effects. One can see that, for small values of the contagion probability as $\gamma=0.1$, the system achieves an absorbing state where $M=1$ and $R=0$. In other words, the radicalization is extinct, and there are only moderated individuals in the population (see Fig. \ref{fig1} (a)). If we increase $\gamma$, for example for $\gamma=0.3$, we observe the coexistence of the two subpopulations $R$ and $M$, i.e., the absorbing state is destroyed. It indicates a phase transition, that will be discussed in more details in the following. For nonzero $\delta$, for example for $\delta=0.1$, we cannot observe such transition: the two subpopulations $M$ and $R$ survive in the long-time limit (see Fig. \ref{fig1} (c) and (d)).

Some of the previous results can be better understanding by simple analytical calculations. Considering the stationary states, the limit $t\to\infty$ in Eqs. \eqref{eq4} or \eqref{eq5} lead to
\begin{equation} \label{eq7}
R=\frac{\delta\,M}{\epsilon-\gamma\,M} 
\end{equation}
\noindent
i.e., a relation between the stationary values $R=R(t\to\infty)$ and $M=M(t\to\infty)$. Considering the normalization condition, Eq. \eqref{eq6}, we obtain a second-order polynomial for $R$, namely $A\,R^{2}+B\,R + C =0$, where
\begin{eqnarray} \label{eq8}
A & = & \gamma \\ \label{eq9}
B & = & \epsilon-\gamma+\delta \\ \label{eq10}
C & = & -\delta
\end{eqnarray}

An interesting limiting case to be analyzed is $\delta=0.0$, i.e., the absence of the pressure exerted by messages spread on social media over the opinion of moderated individuals. In this case, we have $C=0$ in the above second-order polynomial and two simple solutions for $R$. The first solution is given by $R=0$, i.e., the radicals disappear of the population in the long-time limit, and we have $M=1$ through the normalization condition. The second solution is given by $R= 1 - \epsilon/\gamma$, valid for $\gamma>\epsilon$ since it leads to $R>0$. Considering the stationary density of radicals as the order parameter of the model, the last result can be rewritten in the language of critical phenomena as $R \sim (\gamma - \gamma_c)^{\beta}$, where $\gamma_c=\epsilon$ and $\beta=1$. This result suggests a typical active-absorbing phase transition in the directed percolation universality class \cite{javarone_galam,radical_nuno_2023,dickman_book,pla_allan,royal_marcelo}, as observed also for the standard SIS model. This absorbing state was also observed previously, when we discussed the time evolution of the populations of radicals and moderated individuals (see Fig. \ref{fig1}, panel (a)).

Looking for the previously obtained second-order polynomial for $R$, we can see that for $\delta\neq 0.0$ the solution $R=0$ is not valid anymore. In such a case, the solution is given by
\begin{equation} \label{eq11}
R=\frac{1}{2A}\,(-B \pm \sqrt{\Delta}), 
\end{equation}
\noindent
where $\Delta=B^{2} - 4AC$ and $A$, $B$ and $C$ are given by Eqs. \eqref{eq8} - \eqref{eq10}. From such solution, the stationary density of moderated individuals $M$ can be obtained from Eq. \eqref{eq7}.

\begin{figure}[t]
\begin{center}
\vspace{6mm}
\includegraphics[width=0.4\textwidth,angle=0]{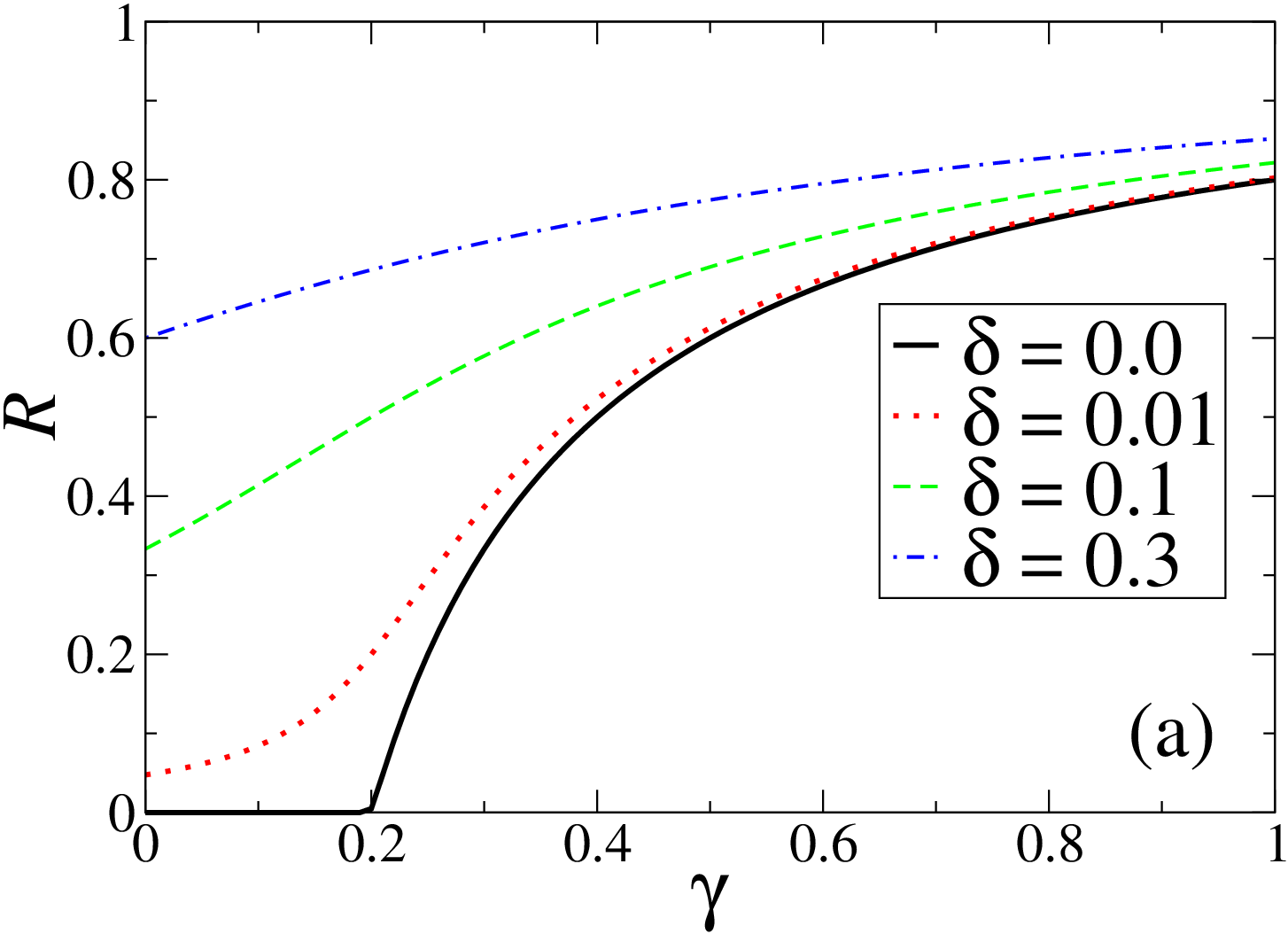}
\hspace{0.2cm}
\includegraphics[width=0.4\textwidth,angle=0]{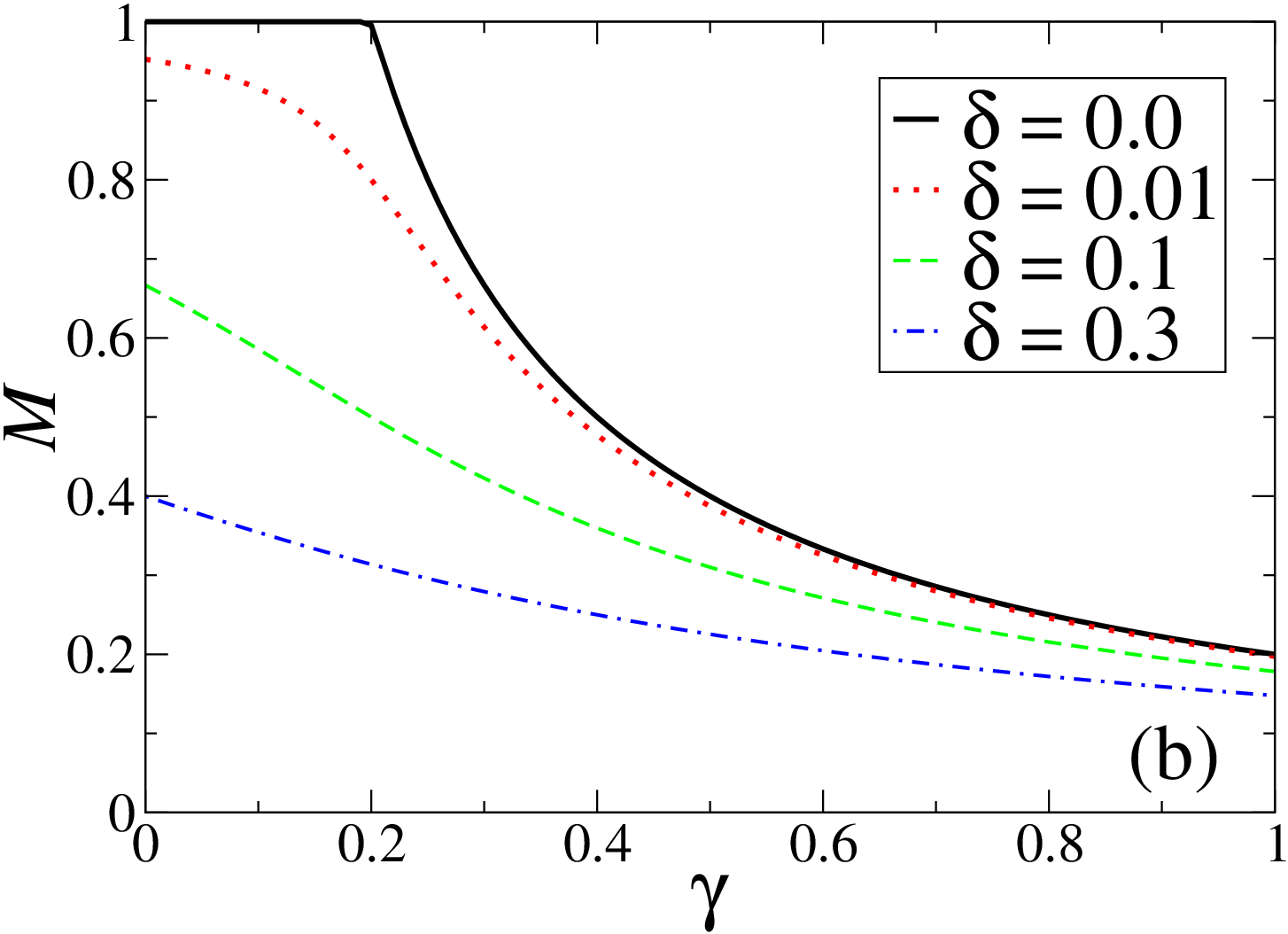}
\end{center}
\caption{(Color online) Stationary densities of radical $R$ (a) and moderated $M$ (b) individuals as functions of the contagion probability $\gamma$. The fixed parameter is $\epsilon=0.2$. We can see that the active-absorbing phase transition is destroyed for nonzero $\delta$, that is a measure of social media effects over moderated individuals.}
\label{fig2}
\end{figure}

To illustrate the previous starionary results, we exhibit in Fig. \ref{fig2} the stationary fractions of radical $R$ (panel (a)) and moderated $M$ individuals (panel (b)) as functions of $\gamma$ for fixed $\epsilon=0.2$ and typical values of $\delta$. The results were obtained by the numerical integration of Eqs. \eqref{eq4} and \eqref{eq5}. As previously obtained analytically, one can observe an active-absorbing phase transition for $\delta=0.0$, i.e., in the absence of the impact of social media on the individuals' opinions. However, even in such case, for sufficient high contagion probability $\gamma$, the radicalism can survive in the population. For $\delta=0.0$, it occurs for $\gamma>\gamma_c=\epsilon$, as above discussed. In the present case of Fig. \ref{fig2}, we have $\gamma_c=0.2$. For $\delta\neq 0.0$, the radicalism survives in the population even for small $\gamma$, which suggests a very important impact of social media effects on the radicalization phenomena. It is indeed a realistic feature of the model, since the social media in Brazil was fundamental to promote the January 8, 2023 extremist attacks in Brazil, as discussed in the Introduction \cite{forbes,chathamhouse}. Indeed, the radicalism spreads fast in the model for increasing $\delta$: for example for $\gamma=0.3$, we have $R\approx 0.33$ for $\delta=0.0$; $R\approx 0.58$ for $\delta=0.1$; and $R\approx 0.72$ for $\delta=0.3$. 

\begin{figure}[t]
\begin{center}
\vspace{6mm}
\includegraphics[width=0.4\textwidth,angle=0]{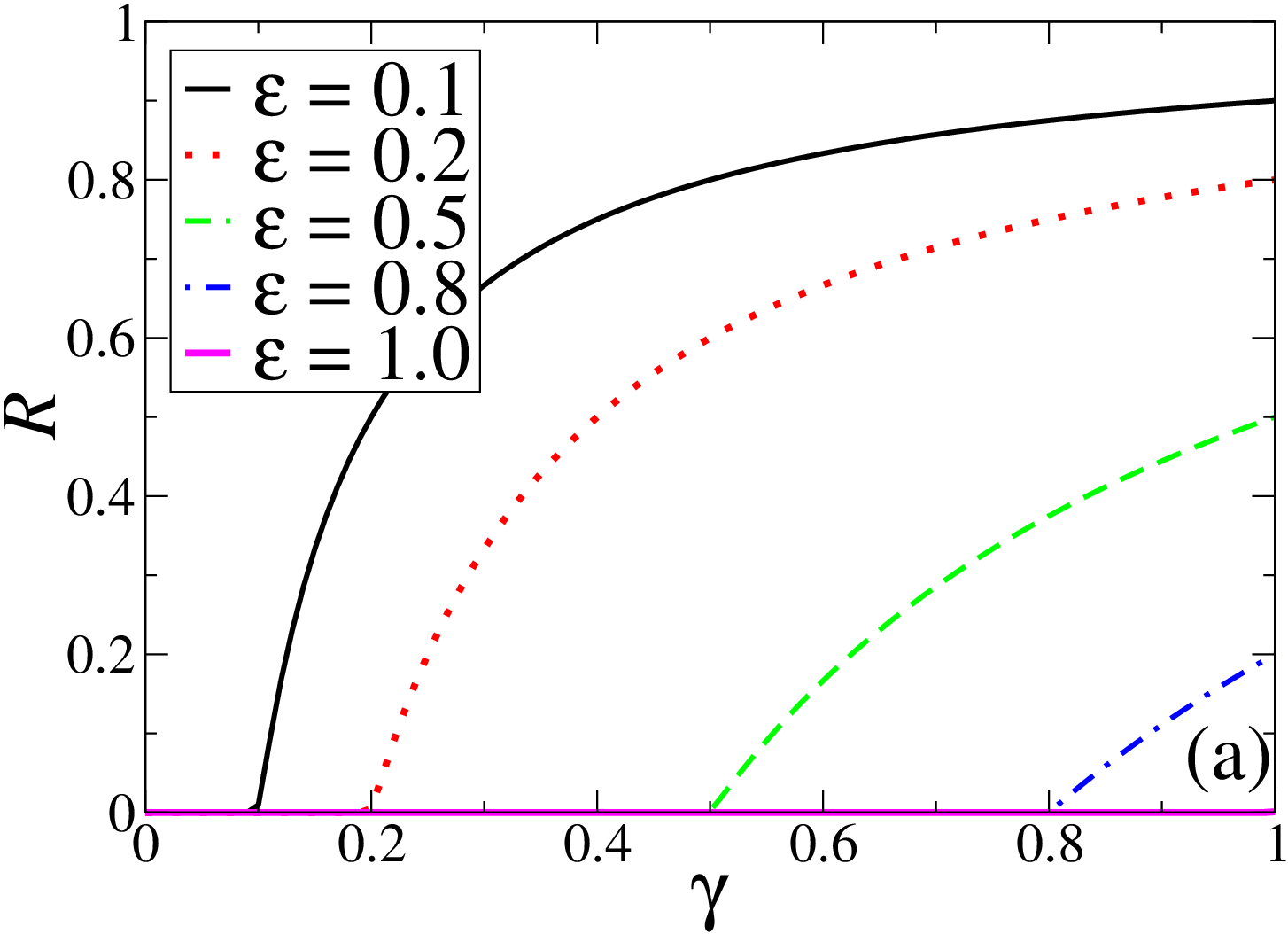}
\hspace{0.2cm}
\includegraphics[width=0.4\textwidth,angle=0]{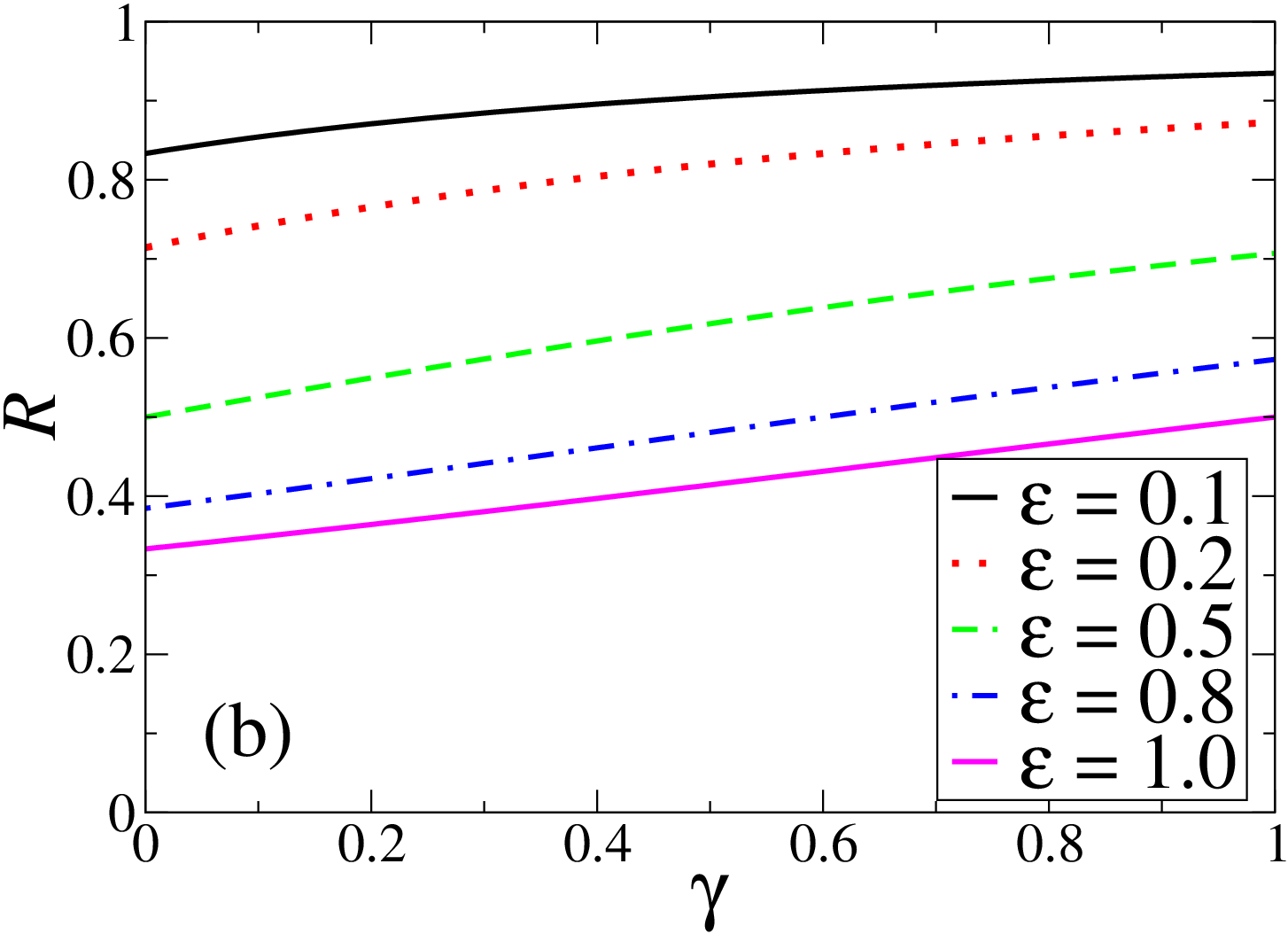}
\end{center}
\caption{(Color online) Stationary density of radicals $R$ as a function of the contagion probability $\gamma$, for typical values of $\epsilon$. The fixed parameters are: (a) $\delta=0.0$ and (b) $\delta=0.5$. We can see that an increase of $\epsilon$, i.e., a higher action of the Federal Supreme Court over extremists, can effectively control the spreading of radicalism, even in the presence of strong social media effects like $\delta=0.5$.}
\label{fig3}
\end{figure}

To understand the impact of the control mechanism, modeled here through the parameter $\epsilon$, we exhibit in Fig. \ref{fig3} the stationary density of radicals $R$ as a function of $\gamma$ for typical values of $\epsilon$. The values of $\delta$ are $\delta=0.0$ (Fig. \ref{fig3} (a)) and $\delta=0.5$ (Fig. \ref{fig3} (b)). As obtained analytically, for $\delta=0.0$ we observe the phase transition for any value $\epsilon<1.0$, at critical points given by $\gamma_c=\epsilon$. However, the density of radicals decreases quickly for increasing values of $\epsilon$. The radicalism can be even extincted for a fixed $\gamma$ and sufficient high values of $\epsilon$. Even in the case $\delta>0.0$ the radicalism can be considerably reduced for increasing values of $\epsilon$. Thus, the model's results show that radicalism can be effectively controlled if we have an active Supreme Court.

\section{Final Remarks}   

\qquad In this work we study a simple contagion model for the spreading of radicalism in an artificial population. Motivated by recent extremism events organized by the supporters of the former far-right Brazilian president Jair Bolsonaro, we propose a SIS-like epidemic model considering two subpopulations, namely moderated and radical agents. The transitions between these two compartments are ruled by probabilities.

Our results suggest that the social media has a fundamental impact on the spreading of extremism in the population. In the absence of such effect, modeled in the present work by a probability $\delta$, the radical population can be extincted in the long-time limit, i.e., in the stationary states of the model (for $\delta=0.0$). The extinction-persistence of the radical population for $\delta=0.0$ is associated with an active-absorbing nonequilibrium phase transition, as the one observed in the standard SIS model. On the other hand, for $\delta\neq 0.0$, i.e., in the presence of social media, the dissemination of messages and fake news in such social media can lead to a considerably increase of the radical population.

The action of the Brazilian Federal Supreme Court was also considered in the model in a simple way, governed by a probability $\epsilon$. It models the deradicalization of such extremist individuals, that can be arrested by the Supreme Court's ministers as occurred in Brazil in January, 2023. Our results suggest that a strong action of the Supreme Court can be effective in control the radicalism in the population.


\section*{Acknowledgments}

The author acknowledges financial support from the Brazilian scientific funding agencies Conselho Nacional de Desenvolvimento Cient\'ifico e Tecnol\'ogico (CNPq, Grant 310893/2020-8) and Funda\c{c}\~ao de Amparo \`a Pesquisa do Estado do Rio de Janeiro (FAPERJ, Grant 203.217/2017).

\bibliographystyle{elsarticle-num-names}

\begin{thebibliography}{00}



\bibitem{social_rmp} 
C. Castellano, S.Fortunato, V. Loreto, \textit{Statistical physics of social dynamics}, Reviews of Modern Physics 81, 591 (2009).

\bibitem{galam_frontiers}
S. Galam, \textit{Physicists, non physical topics, and interdisciplinarity}, Front. Phys. 10:986782 (2022).


\bibitem{sznajd}
K. Sznajd-Weron, J. Sznajd, T. Weron, \textit{A review on the Sznajd model — 20 years after}, Physica A 565, 125537 (2021).


\bibitem{galam_review}
S. Galam, \textit{SOCIOPHYSICS: A REVIEW OF GALAM MODELS}, Int. J. Mod. Phys. C 19, 409-440 (2008).


\bibitem{nuno_pla_2014}
N. Crokidakis, \textit{Phase transition in kinetic exchange opinion models with independence}, Phys. Lett. A 378, 1683 (2014).

\bibitem{nuno_andre_marcelo_pre}
A. L. Oestereich, M. A. Pires, N. Crokidakis, \textit{Three-state opinion dynamics in modular networks}, Phys. Rev. E 100, 032312 (2019).
  

\bibitem{nuno_lucas1}
N. Crokidakis, L. Sigaud, \textit{Modeling the evolution of drinking behavior: A Statistical Physics perspective}, Physica A 570, 125814 (2021).


\bibitem{nuno_lucas2}
N. Crokidakis, L. Sigaud, \textit{Role of inflexible minorities in the evolution of alcohol consumption}, Journal of Statistical Mechanics 093403 (2022).




\bibitem{jin}
F. Jin, Zi-Shan Qian, Yu-Ming Chu, M. ur Rahman, \textit{ON NONLINEAR EVOLUTION MODEL FOR DRINKING BEHAVIOR UNDER CAPUTO-FABRIZIO DERIVATIVE}, Journal of Applied Analysis $\&$ Computation 12(2): 790-806 (2022).



\bibitem{nuno2022}
N. Crokidakis, \textit{A simple mechanism leading to first-order phase transitions in a model of tax evasion}, Int. J. Mod. Phys. C 33, 2250075 (2022).



\bibitem{iglesias}
M. B. Gordon, M. F. Laguna, S. Gon\c{c}alves, J. R. Iglesias, \textit{Adoption of innovations with contrarian agents and repentance}, Physica A 486, 192-205 (2017).



\bibitem{liqing}
Q. Liqing, L. Shuqi, \textit{SVIR rumor spreading model considering individual vigilance awareness and emotion in social networks}, Int. J. Mod. Phys. C 32, 2150120 (2021).







\bibitem{Van_Hiel}
A. V. Hiel, J. V. Assche, T. Haesevoets, D. De Cremer, G. Hodson, \textit{A Radical Vision of Radicalism: Political Cynicism, not Incrementally Stronger Partisan Positions, Explains Political Radicalization}, Political Psychology 43, 3-28 (2022).  


\bibitem{kleinfeld}
R. Kleinfeld, \textit{The Rise of Political Violence in the United States}, Journal of Democracy 32 (4), 160–76 (2021).



\bibitem{times_of_israel}
$https://www.timesofisrael.com/liveblog-january-8-2023/$


\bibitem{renew_group}
$https://www.reneweuropegroup.eu/news/2023-01-18/attack-on-democracy-in-brazil-the-eu-must-propose-new-legislation-to-fight-fascism-and-extremism-online$  


\bibitem{forbes}
$https://www.forbes.com/sites/angelicamarideoliveira/2023/01/13/in-the-aftermath-of-riots-brazil-faces-the-challenge-of-countering-online-radicalization/?sh=44d0bd1633d7$


\bibitem{chathamhouse}
$https://www.chathamhouse.org/2023/01/digital-politics-threatens-democracy-and-must-change$



\bibitem{amit}
S. Amit, L. Barua, A. -Al Kay, \textit{Countering violent extremism using social media and preventing implementable strategies for Bangladesh}, Heliyon 7, e07121 (2021).

  

  
\bibitem{asif}
M. Asif, A. Ishtiaq, H. Ahmad, H. Aljuaid, J. Shah, \textit{Sentiment analysis of extremism in social media from textual information}, Telematics and Informatics 48, 101345 (2020).



\bibitem{kenyon}
J. Kenyon, J. Binder, C. Baker-Beall, \textit{Exploring the role of the Internet in radicalisation and offending of convicted extremists}, Ministry of Justice Analytical Series, Her Majesty's Prison and Probation Service, United Kingdom (2021).  

  
  

\bibitem{brasil_de_fato}
$https://www.brasildefato.com.br/2023/01/10/how-brazilian-politics-reacted-to-bolsonarist-terrorism-in-brasilia$
  

\bibitem{brazilian_report}
$https://brazilian.report/liveblog/2023/01/23/supreme-court-january-8-investigations/$
  

\bibitem{france24}
$https://www.france24.com/en/americas/20230114-brazil-s-supreme-court-agrees-to-investigate-bolsonaro-over-riot$
  
  




\bibitem{santoprete}
M. Santoprete, F. Xu, \textit{Global stability in a mathematical model of de-radicalization}, Physica A 509, 151–161 (2018).


\bibitem{santoprete2}
M. Santoprete, \textit{Countering violent extremism: A mathematical model}, Applied Mathematics and Computation 358, 314–329 (2019).


\bibitem{sooknanan}
J. Sooknanan, D. M. G. Comissiong, \textit{When behaviour turns contagious: the use of deterministic epidemiological models in modeling social contagion phenomena}, Int. J. Dynam. Control 5, 1046–1050 (2017).




\bibitem{nguyen}
V. X. Nguyen, G. Xiao, J. Zhou, G. Li, B. Li, \textit{Bias in social interactions and emergence of extremism in complex social networks}, Chaos 30, 103110 (2020).



\bibitem{javarone_galam}
  S. Galam, M. A. Javarone, \textit{Modeling Radicalization Phenomena in Heterogeneous Populations}, PLoS ONE 11(5): e0155407 (2016).



\bibitem{galam_radicals_2022}
S. Galam, R. R. W. Brooks, \textit{Radicalism: The asymmetric stances of radicals versus conventionals}, Phys. Rev. E 105, 044112 (2022).




\bibitem{radical_nuno_2023}
N. Crokidakis, \textit{Radicalization phenomena: Phase transitions, extinction processes and control of violent activities}, Int. J. Mod. Phys. C (2023), $doi:10.1142/S0129183123501000$.



  



  
\bibitem{prison}
$https://brazilian.report/liveblog/2023/01/09/post-riot-arrests-brasilia/$



\bibitem{dickman_book}
J. Marro, R. Dickman, \textit{Nonequilibrium phase transitions in lattice models} (Cambridge University Press, 2005).


\bibitem{pla_allan}
A. R. Vieira, N. Crokidakis, \textit{Noise-induced absorbing phase transition in a model of opinion formation}, Phys. Lett. A 380, 2632 (2016).   


\bibitem{royal_marcelo}
M. A. Pires, N. Crokidakis, \textit{Double transition in kinetic exchange opinion models with activation dynamics}, Phil. Trans. R. Soc. A 380, 20210164 (2022).



  

  
\end{thebibliography}

\end{document}